\documentclass[aps,twocolumn,superscriptaddress,showpacs,floatfix]{revtex4}

\usepackage{amsmath}
\usepackage{amsbsy}
\usepackage{amssymb}
\usepackage[dvipdfmx]{graphicx}
\usepackage[usenames, dvipsnames]{color} % gestion de différentes couleurs
\usepackage[normalem]{ulem}
\usepackage{subfigure}
\usepackage{verbatim}
\renewcommand\O[1]{\mathrm{O}\left (#1\right )}

\renewcommand{\r}{\vec{r}}

\renewcommand{\k}{\vec{k}}

\newcommand{\kb}{k_{\text{B}}}
\newcommand{\F}{\mathrm{F}}
\newcommand{\Fc}{\vec{\F}_{\rm{c}}}
\newcommand{\Fs}{\vec{\F}_{\rm{s}}}

\newcommand{\rhok}{\hat{\rho}}
\newcommand{\be}{\begin{equation}}
\newcommand{\ee}{\end{equation}}
\newcommand{\br}{\vec{r}}
\newcommand{\bk}{\vec{k}}
\usepackage{cancel}

\def\bse{\begin{subequations}}
\def\ese{\end{subequations}}

\begin{document}

\title{Towards a measurement of the Debye length in very large Magneto-Optical traps}%

\author{J. Barr\'e}
\affiliation{Institut Denis Poisson, Universit\'e d'Orl\'eans, Universit\'e de Tours, CNRS, France,\\
and Institut Universitaire de France}
\author{R. Kaiser}
\author{G. Labeyrie}
\affiliation {Universit\'e C\^ote d'Azur, CNRS, Institut de Physique de Nice, 06560 Valbonne, France;\\}
\author{B. Marcos}
\affiliation{Universit\'e C\^ote d'Azur, CNRS, Laboratoire J.-A. Dieudonn\'e,  06109 Nice, France.}
\author{D. M\'etivier}
\affiliation{Universit\'e C\^ote d'Azur, CNRS, Laboratoire J.-A. Dieudonn\'e,  06109 Nice, France.}
\affiliation{{\rm Present address:} Center for Nonlinear Studies and Theoretical Division T-4 of Los Alamos National Laboratory, NM 87544, USA}

\date{\today{}}

% ----------------------------------------------------------------
\begin{abstract}
We propose different experimental methods to measure the analog of the Debye length in a very large Magneto-Optical Trap, which should characterize the spatial correlations in the atomic cloud. 
An analytical, numerical and experimental study of the response of the atomic cloud to an external modulation potential suggests that this Debye length, if it exists, is significantly larger than what was expected.
\end{abstract}
%
% ----------------------------------------------------------------
%\pacs{} 

\maketitle
 
\section{Introduction}

Magneto Optical Traps (MOTs), first realized in 1987 \cite{raab_trapping_1987}, are still an ubiquitous device to manipulate cold atoms. 
Early studies \cite{walker_sesko_wieman} have shown that when the number of trapped atoms is increased beyond a certain level, the peak density tends to saturate. This unwanted limitation to obtain high spatial densities of laser-cooled atomic samples has been attributed to an effective repulsion between atoms due to multiple scattering of photons. 
A basic model to describe atoms in a large MOT has then emerged, where atoms, beyond the friction and external trapping force, are subjected to two kinds of effective interaction forces: an effective Coulomb repulsion of \cite{walker_sesko_wieman}, which is dominant, and an effective attraction, sometimes called shadow effect, first described in \cite{dalibard1988}. 
Even though the shortcomings of this model are well known (such as a too large optical depth, space dependent trapping parameters \cite{townsend1995phase}, sub-doppler mechanisms \cite{dalibard_laser_1989,kim2004direct}, light assisted collisions \cite{Weiner1999} and radiative escape \cite{Bradley2000, Caires2004} or hyperfine changing collisions \cite{Sesko1989, Lee1996}), its predictions on the size and the shape of the atomic clouds are in reasonable agreement with experiments on very large MOTs \cite{camara_scaling_2014}. \\

It is striking that the above ``standard model" describes MOTs as a kind of analog of a non neutral plasma, as well as an instance of an experimentally controllable system with long range interactions. This has prompted several studies \cite{pruvost_expansion_2000,labeyrie_self-sustained_2006,Pohl2006,mendonca2008collective,mendonca2011laser,tercas_driven_2010,mendonca2012}, aimed at better probing this analogy and its consequences. We note that these long range forces stem from the resonant dipole-dipole coupling between atoms \cite{Courteille2010, Bienaime2011, Chomaz2012, Bienaime2013, Zhu2016, Jenkins2016, Corman2017}, which if interference can be neglected lead to radiation trapping of light in cold atoms \cite{Fioretti1998, Labeyrie2004, Labeyrie2005}. This dipole-dipole coupling is also at the origin of modified radiation pressure on the center of mass \cite{Bienaime2010,Chabe2014} and of optical binding with cold atoms \cite{Maximo2018} as well as of super-subradiance \cite{Guerin2016, Araujo2016, Roof2016}.\\ 

Current technologies now allow for larger and larger MOTs, for which long range interactions become even more important. Hence it becomes feasible to test more quantitatively this plasma analogy. In particular, spatial correlations in plasmas are controlled by a characteristic length, called the Debye length, which depends on charge, density, temperature. A natural question thus arises: is an experimental observation of a Debye length possible in a large MOT?\\
 
In this paper, we propose and analyze three types of experiments to probe spatial correlations in a MOT. We first explain how an analysis of the density profile in the MOT provides an indirect measurement of the Debye length. Then we present a direct measurement by diffraction, and highlight its inherent difficulties: we have not been able to measure spatial correlations this way. Finally, we demonstrate that the cloud's response to an external modulation should also provide an indirect measurement of the Debye length. Our experimental results then show that if the interactions are indeed adequately described by a Coulomb-like interaction, the corresponding Debye length is much larger than what could be expected based on the observed size of the cloud without interaction.

To our knowledge, this is the first attempt to characterize density-density correlations in MOTs. This problem has been tackled in various circumstances for quantum gases (see for instance \cite{Manz2010,Hung2011}); however, in most cases, the density variations of interest were much stronger than those we would like to see in a MOT: a direct imaging of the gas was then often enough to extract the correlations.

In section \ref{sec:model}, we present our experimental set-up, recall the basic features of the "standard model", based on \cite{walker_sesko_wieman}, and discuss the relevant orders of magnitudes.
In section \ref{sec:experiments}, we explain the different options to probe the interactions and correlations inside the cloud: i) analysis of the density profile \ref{sec:density} 
ii) direct diffraction experiments \ref{sec:diffraction} iii) response to an external modulation \ref{sec:modulation}. While method ii) proves to be not viable with current techniques, comparison of analytical results, simulations and experiments for methods i) and iii) suggest that the Debye length in the cloud may be much larger than expected. The last section \ref{sec:conclusion} is devoted to a discussion of these results. Some technical parts are detailed in two appendices.

\section{Experimental setup and standard theoretical model}
\label{sec:model}

\subsection{Experimental setup}
\label{sec:exp_setup}
The experimental apparatus used in this work as been described in detail elsewhere~\cite{camara_scaling_2014}. $^{87}$Rb atoms are loaded in a magneto-optical trap from a dilute room-temperature vapour. The trapping force is obtained by crossing six large laser beams (waist 2.4 cm) at the center of the vacuum chamber, arranged in a two-by-two counter-propagating configuration. These lasers are detuned from the $F = 2 \rightarrow F' = 3$ atomic transition of the D2 line 
by a variable $\delta$, whose value is used  to vary the atom number and size of the cloud. Typically, $\delta$ is  varied from -3$\Gamma$ to -8$\Gamma$, 
where $\Gamma$ is the atomic linewidth. The peak intensity in each beam is 5 mW/cm$^2$. The trapping beams also contain a small proportion (a few $\%$) of ``repumping'' light, tuned close to the $F = 1 \rightarrow F' = 2$ transition. A pair of coils with opposite currents generate the quadrupole magnetic field necessary for trapping. The magnetic field gradient along the axis of the coils is 7.2 G/cm. Due to the large diameter of the trapping beams, the maximal number of trapped atoms is large, up to $10^{11}$. As discussed in the following, this results in a large effective repulsive interaction between atoms mediated by scattered photons. As a consequence the cold atomic cloud is large with a FWHM diameter typically between $10$ and $15$ mm , depending on the value of $\delta$. The temperature of the cloud is of the order 100-200 $\mu$K.

We now describe the various experimental techniques implemented to probe spatial correlations inside the atomic cloud. The results of these experiments and their comparison with theoretical models are presented in section~\ref{sec:experiments}. The first technique simply relies on the analysis of the cloud's density profile. This is achieved by imaging the trapping light scattered by the atoms, known as ``fluorescence'' light, with a CCD camera. However, the spatial distribution of fluorescence light usually does not reflect that of the atomic density, because of multiple scattering~\cite{camara_scaling_2014}. To minimize this effect, we acquire the fluorescence image at a large detuning of $-8\Gamma$. The time sequence is as follows: the MOT is operating at a given detuning $\delta$ (variable), then the detuning is jumped to $-8\Gamma$ for a duration of 10 $\mu$s, during which the image is recorded. During this short time, the atoms move only by a few 10 $\mu$m, which is much smaller than all spatial scales we look for.

The second technique is based on the direct diffraction of a probe beam by the cloud. A weak beam of waist 2.2 mm (much smaller than the cloud's diameter), detuned by several $\Gamma$, is sent through the center of the cloud immediately after the trapping beams are shut down. The transmitted far field intensity distribution is recorder using a CCD camera placed in the focal plane of a lens.

\begin{figure}[!htbp]
\begin{center}
\includegraphics[width=0.48\textwidth]{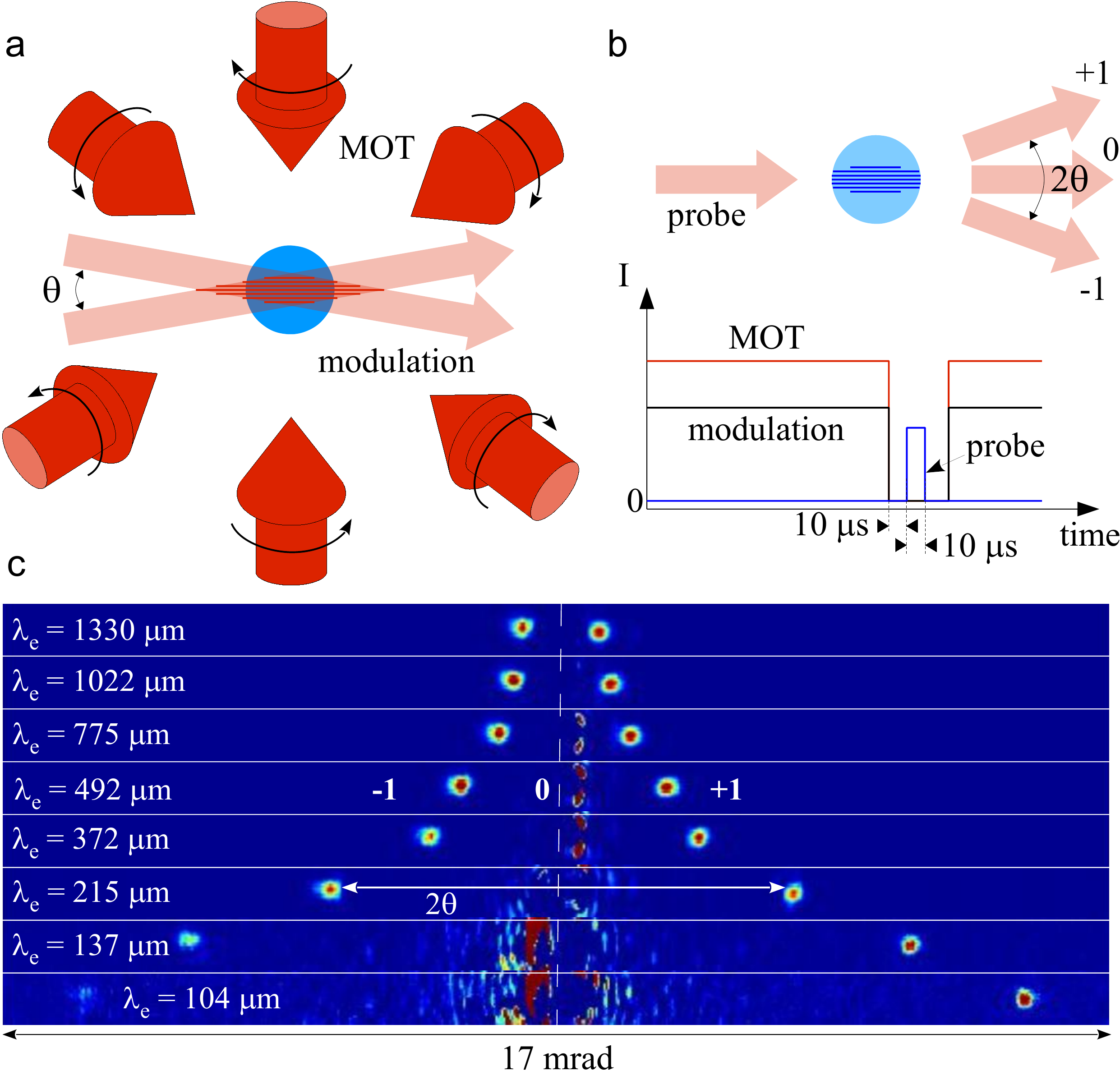}
\caption{Principle of modulation experiment. a: A sinusoidal modulation is applied by crossing two laser beams on the cloud. b: The atoms are released from the MOT and the diffraction grating due to the atomic density modulation is probed. c: Images of the $\pm 1$ diffracted orders versus modulation wavelength $\lambda_e$.}
\label{fig:modulation}
\end{center}
\end{figure}

The third technique relies on the measurement of the cloud's response to an external sinusoidal modulation. Its principle is illustrated in Fig.~\ref{fig:modulation}. A sinusoidal potential is generated by crossing two identical laser beams of waist 2.2 mm and detuning $+20\Gamma$ in the center of the cloud, with an adjustable small angle $\theta$ between them (Fig.\ref{fig:modulation}a). The resulting modulation period is $\lambda_e = \lambda_i/\theta$ where $\lambda_i=780\,$nm is the laser wavelength. The intensity of these beams is chosen low enough such that the associated radiation pressure force doesn't affect the functioning of the MOT (no difference in atom number with and without the modulation beams; the induced density modulation is small, at most a few percent). To measure the response of the cloud (in the form of a density grating), we switch off the MOT laser beams and send the probe beam described before through the modulated part of the cloud. The short delay ($10 \mu$s) between probing and MOT switching off ensures that the initial density modulation is not blurred by the residual atomic motion. The modulated atomic density acts for the probe as a transmission diffraction grating (Fig.\ref{fig:modulation}b). The zeroth and first diffracted orders are recorded by a CCD camera placed in the focal plane of a lens. Fig.\ref{fig:modulation}c 
shows a series of images of the detected diffracted peaks, corresponding to different values of the modulation wavelength $\lambda_e$.  The zeroth order is blocked to avoid saturation of the CCD. As the diffracted light power decreases with $\lambda_e$ (see Fig.\ref{fig:ocp:diff_exp_3}), the display is adjusted for each image of the figure to improve readability.

\subsection{Theoretical methods}
Theoretical descriptions and experimental measurements of density-density correlations are present 
in all fields of condensed matter. We first give below a short introduction to linear response theory and static structure factors, which will play an important role later on (more details can be found for instance in \cite{hansen_theory_2006}).
We define the {\it one-point probability} distribution function $\rho(\br,t)$, usually called {\it density}, as the probability to find a particle at the position $\br$ at time $t$. If the system is statistically homogeneous the density does not depend on the position and time and $\rho(\br,t)=\rho_c$. We define the {\it two-point probability} distribution function  $\rho^{(2)}(\br,\br^{~\prime},t)$ as the probability to find one particle at the position $\br$ and another one at the position $\br^{~\prime}$ at time $t$. $\rho$ and $\rho^{(2)}$ can be expressed as statistical averages of the microscopic one-point and two-point distribution functions:
\begin{eqnarray}
\rho(\br,t) &=& \left\langle \sum_{j=1}^N \delta(\br -\br_j(t))\right\rangle~,\nonumber \\
\rho^{(2)}(\br,\br^{~\prime},t) &=& \left\langle \sum_{j,l=1}^N \delta(\br -\br_j(t))\delta(\br^{~\prime} -\br_l(t))\right\rangle.\nonumber
\end{eqnarray}
It is customary to introduce the function $g$ defined as 
\be
g(\br,\br^{~\prime},t)=\frac{\rho^{(2)}(\br,\br^{~\prime},t)}{\rho(\br,t)\rho(\br^{~\prime},t)}.
\ee
Of central interest in the following will be the structure factor
\begin{equation}
S(\k)=\left <\dfrac{1}{N}\Bigg |\sum_{i}e^{-i\vec{k}\cdot \r_i}\Bigg |^2\right >,\label{eq:structure_factor}
\end{equation}
because it is directly related to the observed diffracted intensity in a diffraction experiment. Both $g$ and $S$ contain information on the density correlations.

If the system is statistically homogeneous, $g(\br,\br^{~\prime},t)$ depends only on $\br-\br^{~\prime}$; if in addition it is statistically isotropic, 
$g$ depends only on $|\br-\br^{~\prime}|$, and will be written $g(r,t)$.
In this case, calling $\rho_c$ the constant density, we have
\begin{eqnarray}
S(\bk)&=&1+\rho_c \int g(r)e^{-i\bk\cdot \br d\br} \nonumber \\
&=& 1+N\delta(\bk) +\rho_c \int [g(r)-1]e^{-i\bk\cdot\br}d\br.
\label{sf}
\end{eqnarray}

We now introduce the linear response theory, which describes the response of the system to a small external perturbation. Consider an uniform system of density $\rho_c$ exposed to a weak external potential $\delta \phi(\br)$. Linear response theory asserts that the density perturbation $\delta \rho$ created by $\delta \phi$ 
is~\cite{hansen_theory_2006}
\be
\label{density-ft}
\delta\hat\rho(\bk)=-\beta \rho_c [S(\bk)-N\delta(\bk)] \delta\hat\phi(\bk).
\ee
We will give an approximate theoretical expression for $S(\bk)$ in a MOT in section \ref{section-analysis}, and use these results in section \ref{sec:modulation}.

\subsection{Model}
In the standard Doppler model, all forces on atoms inside a MOT stem from the radiation pressure exerted by the almost resonant photons. Over long enough time scales, the scattering of many photons produces an average force on the atomic cloud, which may be decomposed as: velocity trapping (ie friction), spatial trapping, attractive shadow effect, and repulsion due to multiple scattering. The first two are single atom effects, the last two are effective interactions between atoms.   
The friction force $F_{\rm dop}$ is due to Doppler cooling. Linearizing for small velocities, it reads
\begin{equation}
\vec{F}_{\rm dop} \simeq -m \gamma \vec{v},
\end{equation}
with
\[
\gamma=\dfrac{I_0}{I_s}\dfrac{8\hbar k_{\rm Las}^2}{m}\frac{-\bar{\delta}}{\left (1+4\bar{\delta}^2 \right )^2},
\] 
where $I_0,k_{\rm Las},\bar{\delta}=\delta/\Gamma$ are respectively the laser intensity, wave number and scaled detuning, $I_s$ is the saturation intensity, and $m$ the atomic mass.
This expression assumes a small saturation parameter. $\gamma$ is positive (actual friction) when the lasers are red detuned ($\delta<0$).

The trapping force $F_{\rm trap}$ is created by the magnetic field gradient. We will consider a linear approximation to this force:
\begin{equation}
\vec{F}_{\rm trap} \simeq -m\omega_x^2  x \vec{e}_x-m\omega_y^2  y \vec{e}_y-m\omega_z^2  z \vec{e}_z.
\end{equation}
The antihelmhotz configuration of the coils induces a non isotropic trap, with $\omega_y^2=\omega_z^2=\frac12 \omega_x^2$. Nevertheless via laser intensity compensations it is possible to obtain a spherical cloud, hence we will use in our modelling $\omega_y=\omega_z=\omega_x=\omega_0$.

The shadow effect, first studied in \cite{dalibard1988}, results from the absorptions of lasers by atoms with cross section $\sigma_L$ in the cloud. 
This force is attractive, and in the small optical depth regime, its divergence is proportional to the density $\rho$:
\begin{equation}
\vec{\nabla}\cdot\Fs=-6I_0\dfrac{\sigma_L^2}{c}\rho(x,y,z),
\label{eq:ocp:poisson_shadow}
\end{equation}
where $c$ is the speed of light. Note however that $\Fs$ does not derive from a potential.

The repulsive force \cite{walker_sesko_wieman} is due to multiple scattering of photons. If the optical depth is small, very few photons are scattered more than twice, and
the effect of multiple scattering can be approximated as an effective Coulomb repulsion
\begin{equation}
\Fc(\r)=3I_0\dfrac{\sigma_L \sigma_R }{2\pi c}\dfrac{\r}{r^3},
\end{equation}
where $\sigma_R$ is the atomic cross section for scattered photons. The divergence of the force is 
\[
\vec{\nabla}\cdot\Fc= 6I_0\dfrac{\sigma_L\sigma_R}{c}\rho(x,y,z). 
\]
The scattered photons actually have complex spectral and polarization properties, and $\sigma_R$ should rather be understood as an averaged quantity. In all experiments, $\sigma_R>\sigma_L$, with the consequence that the repulsion dominates over the attractive shadow effect. Since repulsion and attraction both have a divergence proportional to the local density, the shadow effect is often considered as a mere renormalization of the repulsive force; note that this involves a further approximation, because the forces are not proportional, even though their divergences are.

Finally, the spontaneous emission of photons acts as a random noise on the atoms, which induces at the macroscopic level a velocity diffusion.
In our experiments, the atomic dynamics is typically overdamped: the velocity damping time is much shorter than the position damping time. The velocity distribution then quickly relaxes to an approximate gaussian, with temperature $T$, and the density  $\rho(\r,t)$ is described by the Smoluchowsky equation (which is a simplified version of the Fokker-Planck equation in \cite{romain_phase-space_2011}): 
\begin{equation}
\partial_t \rho(\r,t)=\vec{\nabla}\cdot\left (\omega_0^2\r \rho-\dfrac{1}{m}(\Fc+\Fs)[\rho]\rho+\dfrac{\kb T}{m}\vec{\nabla}\rho \right ),
\label{eq:smolu}
\end{equation}
with a Poisson equation for the force
\begin{equation}
\vec{\nabla}\cdot (\Fc+\Fs)= C \rho~{\rm with}~C=6I_0\dfrac{\sigma_L(\sigma_R-\sigma_L)}{c}.
\label{eq:vfp_poisson_coulomb}
\end{equation}
Note finally that in this simplified framework the total force $\Fc+\Fs$ has the same divergence as an effective Coulomb force
\begin{equation}
\tilde{\vec{F}}_c(\r)= \dfrac{C}{4\pi}\dfrac{\r}{r^3}.  
\label{eq:equiv_coulomb}
\end{equation}

\subsection{Analysis of the model}
\label{section-analysis}
The above model describes a large MOT as a collection of particles in a harmonic trap, and the dominant interacting force is a Coulomb-like repulsion.
This clearly suggests an analogy with non neutral plasmas, where trapped electrons interact through real Coulomb forces; for a detailed review, see \cite{dubin_trapped_1999}.
The analogy is not perfect: for instance the non potential part of the shadow effect is neglected, the friction and diffusion in a MOT are much stronger than in a non neutral plasma, and the typical optical depth in an experiment is not very small. Nevertheless, it is a basic model to analyze MOT physics, and has been used recently to predict new plasma related phenomena in MOTs (see for instance  \cite{mendonca2008collective,terccas2013polytropic}). 

\paragraph{Temperature and repulsion dominated regimes}
When the repulsion force is negligible, the trapping force is balanced by the temperature. The cloud has then a gaussian shape, with atomic density
\begin{equation}
\label{def:lg}
\rho(\vec{r}) = \frac{N}{(2\pi l_g^2)^{3/2}}e^{-\frac{\vec{r}^2}{2l_g}}~,~{\rm with}~l_g = \left(\frac{k_BT}{m\omega_0^2}\right)^{1/2},
\end{equation}
where $N$ is the total number of trapped atoms. In the following, $l_g$ will be called the "gaussian length". 
For typical MOT parameters, one has as an order of magnitude $l_g \sim 200\mu m$. 
Increasing $N$, the repulsion increases, and the system enters the repulsion dominated regime, where the trapping force is balanced by the repulsion. Theory then predicts a spherical cloud with constant density $\rho_c$, and step-like boundaries smoothed over the same length scale 
  $l_g$ defined in Eq.~\eqref{def:lg} \cite{dubin_trapped_1999}; the radius of the cloud at zero temperature is denoted by $L$, and we have the expressions
\begin{equation}
\label{eq:rhoc}
\rho_c =\frac{3m\omega_0^2}{C}=\frac{3m\omega_0^2 c}{6I_0\sigma_L(\sigma_R-\sigma_L)}~,~L = \left(\dfrac{4N}{3\pi\rho_c}\right)^{1/3}.
\end{equation}
The cross over between temperature and repulsion dominated regimes is for $l_g \sim L$.
Experimentally, sizes of order $L \sim 1$\,cm can be reached (see section \ref{sec:exp_setup}), which should be well into the repulsion dominated regime. Note that the repulsion dominated regime is not as straightforward to analyze when the trap anisotropy and shadow effect are taken into account, see \cite{romain_measuring_2014}.
 
\paragraph{Plasma coupling parameter and Debye length.}
To quantify the relative effect of kinetic energy and Coulomb repulsion, it is customary for plasmas to define the ``plasma coupling parameter" $\Gamma_p$, which  
is the ratio of the typical potential energy created by a neighboring charge by the typical kinetic energy. For a MOT in the repulsion dominated regime, denoting $a=(3\rho_c/4\pi)^{-1/3}$ a measure of the typical interparticle distance, we have the expression 
\begin{equation}
\Gamma_p=\dfrac{C/(4\pi a)}{ k_{\rm B}T}=\dfrac{a^2}{l_g^2}    
\label{eq:ocp:gamma}
\end{equation}
where we have used \eqref{eq:rhoc}, and we recall that $l_g=(k_BT/m\omega_0^2)^{1/2}$ is the "gaussian length". Using typical experimental values $l_g = 200\mu m$, and an atomic density $\rho=10^{11} {\rm cm}^{-3}$, this yields $\Gamma_p \sim 10^{-4}$. A plasma experiences a phase transition from liquid phase to solid phase at $\Gamma_p\simeq 175$, and is considered in a gas-like phase as soon as $\Gamma_p<1$. The typical value for a MOT experiment is hence very small, well into the gas phase, and the expected correlations are weak. In this regime, and assuming the MOT shape is dominated by repulsion, so that the density in the central region is approximately constant, Debye-H\"{u}ckel theory can be applied. We give now a short account of this theory. Choosing the origin of coordinates as the position of an atom, the density distribution is given by the Boltzmann factor
\be
\label{d-h-density}
\rho(\br)=\rho_c e^{-\frac{\psi(\br)}{k_BT}},
\ee
where $\psi(\br)$ is the {\em average} potential around $\br=0$. Using the
Poisson equation it is possible to find -- self-consistently -- the average potential:
\be
\label{d-h-poisson}
\nabla^2 \psi(\br)=-C\left[\delta(\br)-\rho_c +\rho_c e^{-\frac{\psi(\br)}{k_BT}}\right],
\ee
where the first term on the r.h.s. represents the point charge of the atom. Using the hypothesis that $\Gamma_p\ll1$, the Poisson equation can be simplified: 
\be
\label{d-h-poisson-s}
\left[\nabla^2 -\kappa_D^2\right]\psi(r)=-C\delta(r),
\ee
where $\kappa_D=\lambda_D^{-1}$ and 
\be
\lambda_D=\left(\frac{k_B T}{\rho_c C}\right)^{1/2}.
\label{eq:lambdaD}
\ee
It is simple to
show that the
solution of Eq. \eqref{d-h-poisson-s} is 
\be
\label{d-h-effective}
\psi(r)=\frac{e^{-r/\lambda_D}}{r}, 
\ee
which yields for the pair correlation function \cite{hansen_theory_2006}
\begin{equation}
g(r) = \exp\left(-a\frac{\Gamma_p}{r}e^{-r/\lambda_D}\right).
\end{equation}
This expression assumes isotropy: this is why the correlation depends only on one distance $r$. Note that isotropy is certainly not exactly true for a MOT. $g$ vanishes for small $r$, which is a manifestation of the strong repulsion, and tends to $1$ for $r\gg \lambda_D$: correlations disappear in this limit.
The excluded volume effect kicks in at very small scales, of order $a\Gamma_p$; at larger scales, the above expression can be replaced by:
\begin{equation}
g(r) \simeq 1-\frac{a\Gamma_p}{r}e^{-r/\lambda_D}~.
\label{eq:pair_correlation}
\end{equation}
From this expression we can compute the structure factor \eqref{sf}:
\be
\label{ocp-sf}
S(k)=N\delta(\bk)+ \frac{k^2}{k^2+\kappa_D^2}.
\ee
For weak plasma parameter $\Gamma_p\to 0$, particles are uncorrelated and Poisson distributed;  there is no characteristic correlation length, $\lambda_D\to\infty$ and 
the structure factor is $S(k)= N\delta(k)+1.$

Inserting in \eqref{eq:lambdaD} the expression for $\rho_c$ \eqref{eq:rhoc}, one obtains the expression $\lambda_D = l_g/\sqrt{3}$, and the rough order of magnitude $\lambda_D \sim 100 \mu m$. Using this and the estimated $\Gamma_p$ in \eqref{eq:pair_correlation}, we see that the correlations are indeed very small over length scales of order $\lambda_D$.

\subsection{Simulations of the "Coulomb model"}
\label{sec:simulations}
We will use in section \ref{sec:experiments} numerical simulations to compare the theory with the experiments. We describe here these simulations.

We use Coulomb Molecular Dynamics (MD) simulations, with
typically $N=16384$ particles in an harmonic trap interacting through Coulombian interactions (without shadow effect), with friction and velocity diffusion. We use a second order Leap-Frog scheme (see e.g. \cite{yoshida_1990}); the interaction force is implemented in parallel on a GPU. We are not interested in dynamical effects, hence in all cases the simulation is run until the stationary state is reached. 

The number of simulated particles is much smaller than the actual number of atoms, which is about $10^{11}$. One simulated particle thus represents many physical atoms, and its mass and effective charge are scaled accordingly. 
The price to pay is that the interparticle distance, and hence the plasma parameter $\Gamma_p$, is much larger in the simulations than in the experiments, see the expression \eqref{eq:ocp:gamma}. However, all simulations remain safely in the gas-like phase $\Gamma_p\ll 1$; in other words the larger interparticle distance should not modify the density profile nor prevent the observation of $\lambda_D$ in simulations. We can have $\Gamma_p$ as low as $\sim 10^{-2}$ in simulations while having $\lambda_D/L\sim 0.1$; for larger $\lambda_D/L$, $\Gamma_p$ may be much smaller, see Fig.~\ref{fig:ocp:dxx} for $\delta/\Gamma=-8$.

\subsection{Experimental probes of the ``Coulomb'' model}

Following \cite{walker_sesko_wieman}, describing the optical forces induced by multiple scattering as an effective Coulomb repulsion is a standard procedure since the early 90s. In particular, it satisfactorily explains the important observation that the atomic density in a MOT has an upper limit (preventing for instance the initially sought Bose-Einstein condensation). It also predicts a size scaling $L\sim N^{\sim 1/3}$, which is observed with reasonable precision in the experiments~\cite{Sesko1991,camara_scaling_2014,gattobigio_scaling_2010,gattobigio_manipulation_2008}. However other mechanisms can lead to an upper bound on the density, such as light assisted collisions or other short range interactions \cite{Anderson1994,Weiner1999,Caires2004}. 
Besides the bounded density and size scaling, there are experiments that are consistent with a Coulomb type repulsion: 
\begin{itemize}
\item A Coulomb explosion in a viscous medium has been observed by measuring the expansion speed of a cold atomic cloud in optical molasses: ~\cite{pruvost_expansion_2000,pruvost_coulomb-like_2012}. The result shows a good agreement with what is predicted for a similar Coulomb gas. 
\item Self-sustained oscillations of a MOT have been reported in~\cite{labeyrie_self-sustained_2006}. The model used to explain the experimental observations assume a cloud with a size increasing with the atom number. This is again consistent with a Coulomb type repulsion but remains a indirect test of these forces.
\end{itemize}
All these experiments rely on identifying macroscopic effects of the repulsive force, and microscopic effects such as the building of correlations in the cloud have not been directly observed. This is our goal in the following.

\section{Looking for correlations in experiments}
\label{sec:experiments}

In order to measure directly or indirectly the interaction induced correlations in the atomic cloud, we have performed three types of experiments, which rely on:
i) an analysis of the density profile, ii) a direct measurement of correlations by diffraction iii) an analysis of the cloud's response to an externally modulated perturbation.
This section gathers our results. 
 
\subsection{Analysis of the density profile} 
\label{sec:density}
From the theoretical analysis presented in the previous section, we know that our basic model \eqref{eq:smolu} relates the Debye length $\lambda_D$, which controls the correlations, to the ``gaussian length'' $l_g$, which controls the tails of the density profile: $\lambda_D =l_g/\sqrt{3}$. Fitting the experimental density profile may then provide information on the Debye length. We recall that this is an indirect method and only serves a a guide for a more reliable estimation of the Debye length.

\begin{figure}[!htbp]
\begin{center}\includegraphics[width=0.45\textwidth]{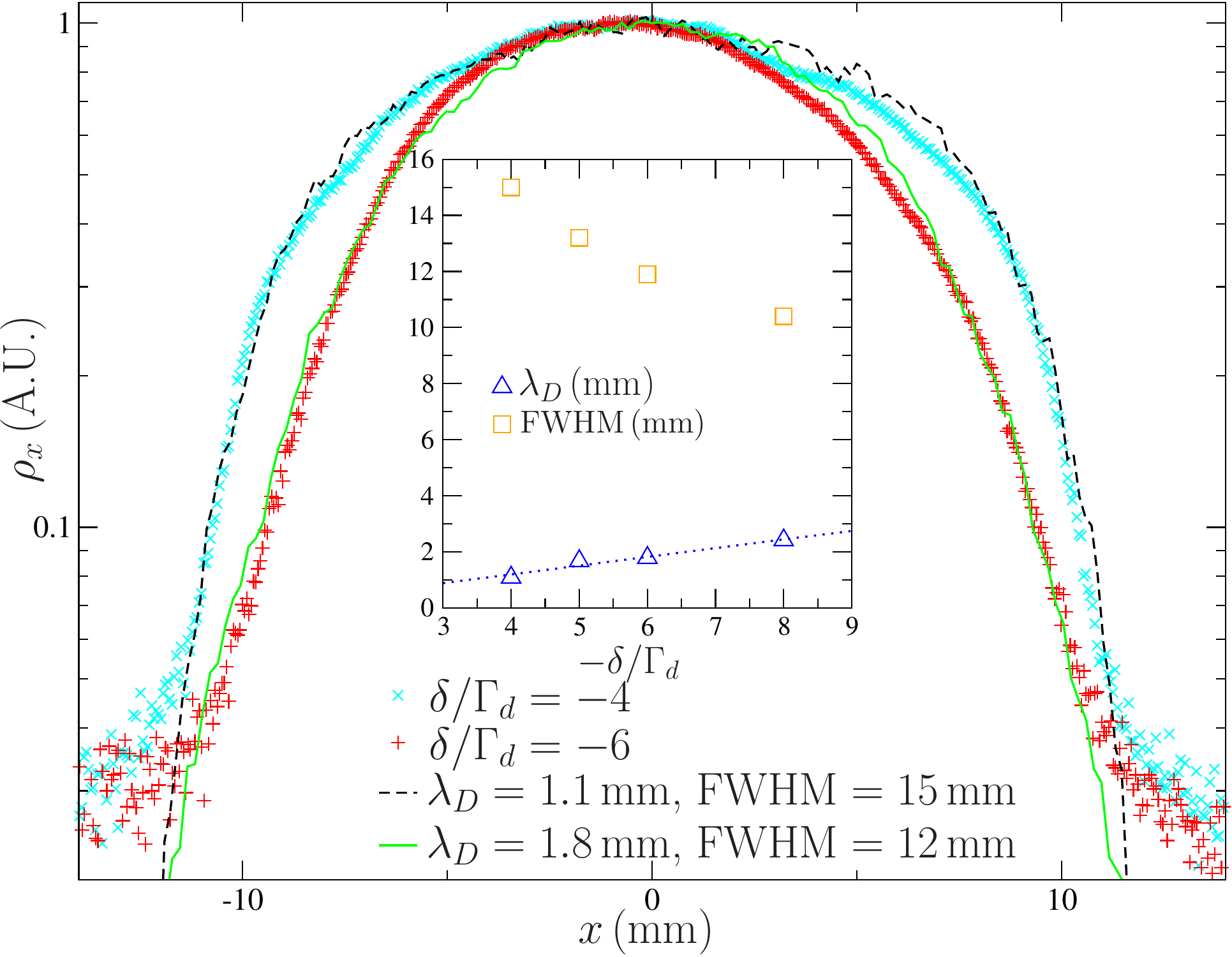}
\caption{Density $\rho_x(x)$ obtained by fluorescence for $-\delta/\Gamma=4,6$ compared with MD simulation of a trapped Coulomb gas, using $N=16384$ particles. The inset shows the extrapolated Debye length $\lambda_D$ and the cloud FWHM diameter. (The density plots for $-\delta/\Gamma=5,8$ are not shown here). The simulated plasma parameter ranges from $\Gamma_p\simeq 4\cdot 10^{-2}$ for $\delta/\Gamma=-4$ to $\Gamma_p\simeq 5 \cdot 10^{-5}$ for $\delta/\Gamma=-8$. For all experiments, the number of trapped atoms is of the order of $10^{11}$.
} 
\label{fig:ocp:dxx}
\end{center}
\end{figure}

The experimental data obtained by fluorescence \cite{camara_scaling_2014} is 
two dimensional, since the density is integrated over one direction (called $z$ below) hence, we cannot see directly $\rho(r)$ but an integrated quantity; selecting the central part $y\in[-\epsilon,\epsilon]$, where $\epsilon$ is about $10\%$ of cloud's width, we obtain the observed density along the $x$ direction:
\[
\rho_x(x)=\int_{-\infty}^{\infty}\mathrm{d} z\int_{-\epsilon}^\epsilon\mathrm{d} y~\rho(x,y,z), 
\]
Figure \ref{fig:ocp:dxx} shows, for two values of the detuning $\delta$, this partially integrated experimental density profile $\rho_x$.

We now compare these profiles with numerical simulations, see subsection \ref{sec:simulations}. 
We choose the simulation parameters by fixing the radius at zero temperature $L$ and the Debye length $\lambda_D$. 
We obtain from the simulations density profiles that depend on $L$ and $\lambda_D$, which we fit to the experimental data. The numbers of simulated particles is much smaller than the actual number of atoms, but simulations are still in the $\Gamma_p\ll 1$ regime, which allows a meaningful fit of the density profile, see \ref{sec:simulations}.
Figure \ref{fig:ocp:dxx} shows that the fits are reasonably good, and allow to extract a value for $\lambda_D$ and $L$, or, equivalently, for $\lambda_D$ and the FWHM.
These results suggest a value for the Debye length in the $1-2$mm range, much larger than what was expected on the basis of the experiments in the temperature dominated regime, see section~\ref{sec:model}. 
However, this method is very model dependent: one could imagine other physical mechanisms or interaction forces producing similar density profiles. 
To overcome this difficulty, we need methods able to probe more directly the interactions and correlations inside the cloud. This is the goal of Sections
\ref{sec:diffraction} and \ref{sec:modulation}.

\subsection{Direct probing of correlations by diffraction}
\label{sec:diffraction}

\begin{figure}[!htbp]
\begin{center}
\includegraphics[width=0.48\textwidth]{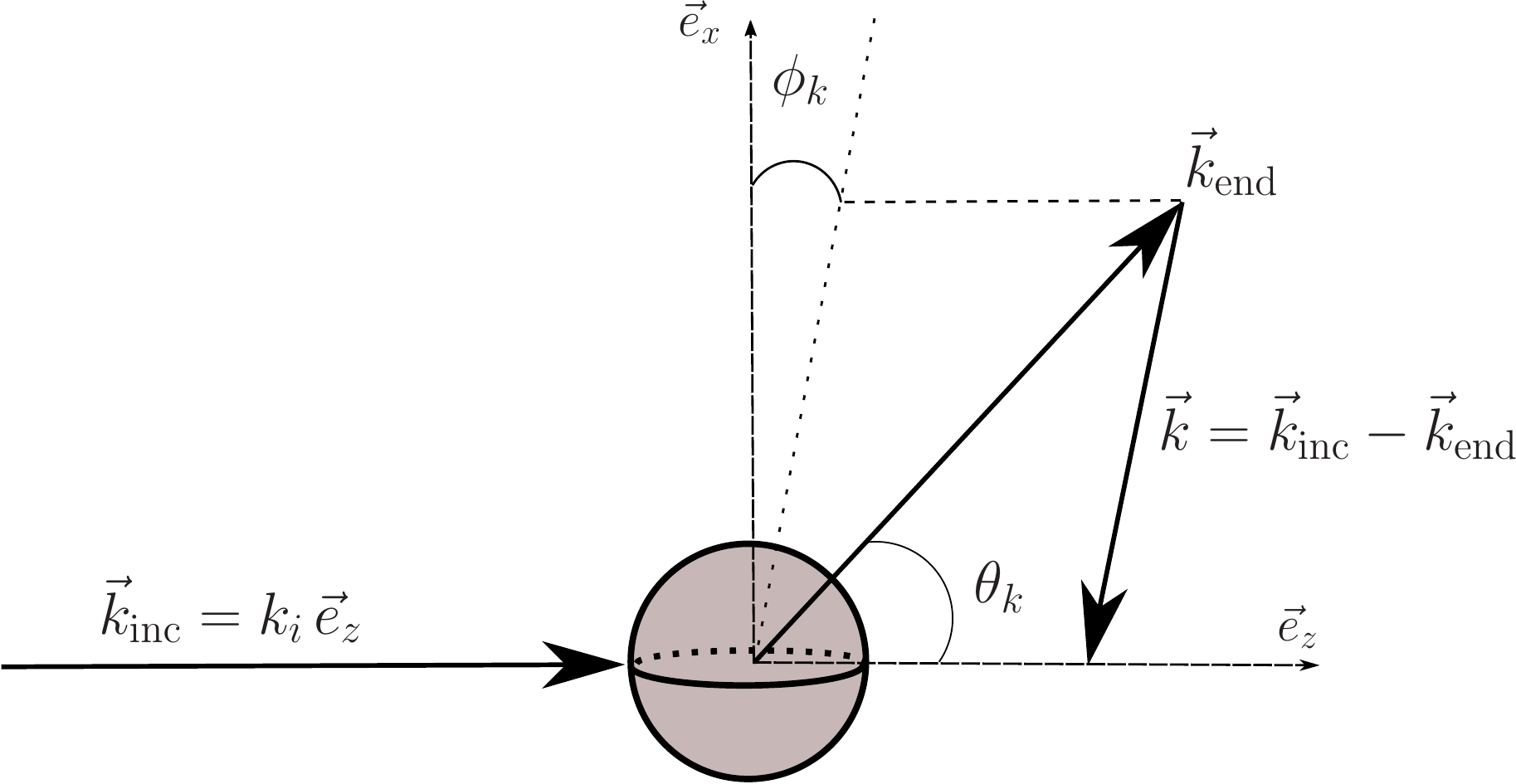}
\caption{Sketch of an incident beam $\k_{\text{inc}}$ diffracted on an atom in direction $\k_{\text{end}}$ corresponding to angles $\theta_k$ and $\phi_k$. We define and show the vector $\k=\k_{\text{inc}}-\k_{\text{end}}$.}
\label{fig:angles}
\end{center}
\end{figure}

An alternative method to probe spatial correlations of particles and thus access the Debye length is by directly probing two-body correlations via a diffraction experiment: an additional detuned laser beam is sent through the cloud, and the diffracted intensity $I$ is recorded. For an incident plane wave, $I$ is proportional to the structure factor $S(\bk)$ given by \eqref{eq:structure_factor}, 
where $\k=\k_{\text{inc}}-\k_{\text{end}}$ is the difference between the incident wavevector $\k_{\text{inc}}=k_i \vec{e}_z$ and the diffracted one $\vec{k}_{\text{end}}=k_i(\cos\phi_k\sin\theta_k,\sin\phi_k\sin\theta_k,\cos\theta_k)$; this assumes elastic scattering, see figure \ref{fig:angles} (see \cite{hansen_theory_2006} for a reference).

\begin{figure}[!htbp]
\begin{center}
\includegraphics[width=0.48\textwidth]{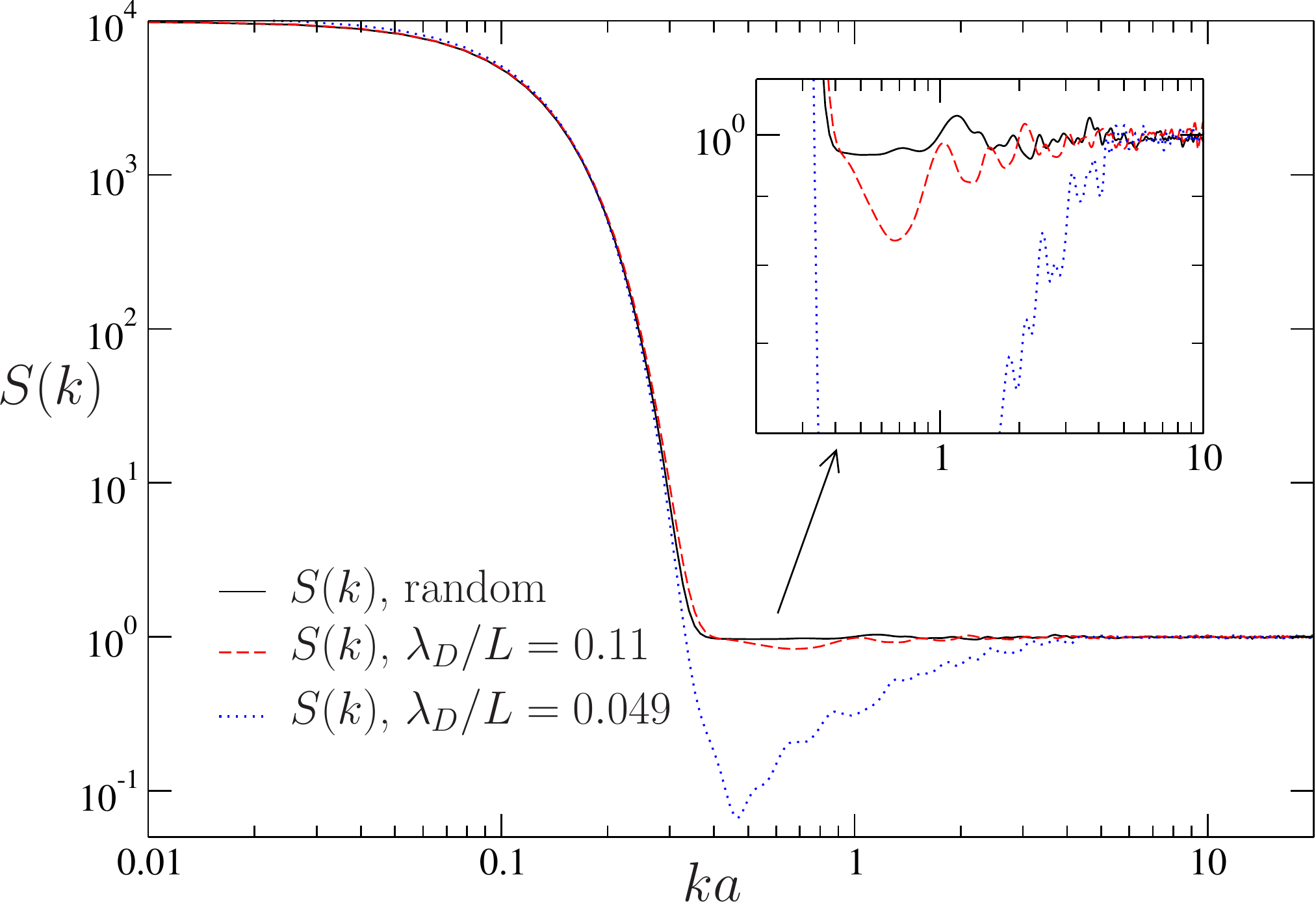}
\caption{MD simulations with $N=16384$ particles of the structure factor $S(k)$, averaged over all $\k$ such that $|\k|=k$. The horizontal axis is adimensionalized by the mean interparticle distance $a$, which is in the simulation $a/L=0.039$. For the dashed red curve $\Gamma_p\simeq 0.043$ with the same ratio $\lambda_D/L$ than the black dashed fit in Fig.~\ref{fig:ocp:dxx}, for the dotted blue curve $\Gamma_p\simeq 0.215$ (these values for the plasma parameter are much higher than expected in the atomic cloud; smaller, more realistic, values are difficult to reach numerically while keeping a small $\lambda_D/L$). The waist of the gaussian probe beam is $w\simeq 0.76L$. The black curve corresponds to randomly distributed particles with the same average density: the two-body correlation obviously vanishes in this case, and accordingly, the characteristic dip is absent.}
\label{fig:ocp:structure}
\end{center}
\end{figure}

We then have
\begin{equation}
k=|\k|=2k_i\sin(\theta_k/2).
\end{equation}
In an isotropic homogeneous infinite medium the theoretical structure factor would be given by \eqref{ocp-sf}.
In the actual experiment, the structure factor \eqref{ocp-sf} is modified at small $k$ either by the finite size of the cloud, or by the finite waist of the probe beam, whichever is smaller: the $\delta$ function is replaced by a central peak which simply reflects the Fourier transform of the density profile or of the beam profile. 
Figure~\ref{fig:ocp:structure} shows an example of $S(k)$ for an MD simulation of a trapped Coulomb cloud, with a gaussian probe beam smaller than the cloud:
\begin{itemize}
%\item The main peak $S(k=0)=N$ corresponds to the %unscattered radiation.
\item For small $k\sim 1/L$, there is a large smooth peak, corresponding to the Fourier transform of the probe beam's profile.
\item For large $k$, the structure factor tends to 1 (this is clear from \eqref{sf}).
\item For intermediate $k\sim 1/\lambda_D$, there is a small dip which is the manifestation of the Debye length. It is deeper when the temperature is smaller, since correlations are stronger. It disappears for large temperature (the black curve in Fig.~\ref{fig:ocp:structure} formally corresponds to an infinite temperature). For values of $\lambda_D/L$ compatible with Fig.\ref{fig:ocp:dxx} (red dashed curve), the dip is barely visible in the simulations.
\end{itemize}
Unfortunately, for $\lambda_D/L$ in the range suggested by 
Sect.~\ref{sec:density}, it is difficult to disentangle the small dip, signature of the Debye length, from the tails of the central peak. Furthermore the ratio dip amplitude / central peak height scales as $1/N_{\rm diff}$, where $N_{\rm diff}$ is the number of diffracting atoms.

\subsection{Response to an external modulation}
\label{sec:modulation}

\subsubsection{Theoretical analysis: Bragg and Raman-Nath regimes}
\label{par:theory}
Since a direct measure of correlations inside the cloud is currently not accessible, we have studied indirectly the effect of these correlations, by analyzing the response
to an external force. The experimental procedure has been described in section \ref{sec:exp_setup}. As we will see below, this response is related to the interactions inside the cloud.

The static modulation potential in the direction $\vec{e}_x$, with amplitude $A$, reads:
\begin{equation}
\phi_{\rm{ext}}(x)=A\sin(k_e x).
\label{eq:modulation}
\end{equation}
Experimentally, the depth of the modulation potential was chosen so that the density modulation never exceeded 10\%; hence we limit ourselves to a linear response computation. We are interested in the diffraction profile, which is proportional to the structure factor $S(\vec{k})$. The location of the diffracted peak is given by the modulation wave vector $k_e$, and the experimentally measured quantity is the integrated diffracted power around $k_e$, denoted $R(k_e)$. The detailed computations are in the appendix, we report here the results. 
The main features are:\\
i) There is a cross-over between the Bragg regime at small modulation wavelength $\lambda_e< \lambda_e^{(c)}$, or $k_e> k_e^{(c)}$, and the Raman-Nath regime at large modulation wavelength $\lambda_e> \lambda_e^{(c)}$, or $k_e< k_e^{(c)}$. We have
\begin{equation}
\lambda_e^{(c)}=2\pi\sqrt{\dfrac{L}{2k_i}}=\sqrt{\pi L\lambda_i}\quad\text{or}\quad k_e^{(c)}=\sqrt{\dfrac{2k_i}{L}}.
\label{eq:ocp:criteria}
\end{equation}
In the Bragg regime, the response is dominated by the longitudinal density profile, whereas in the Raman-Nath regime, the response is dominated by the effect of the interactions inside the cloud: the latter is then of most interest to us. For our experimental conditions, the cross over is around $\lambda_e^{(c)}=120\mu$m.\\
ii) We obtain (see appendix) the approximate expression for the integrated diffracted power:
\begin{equation}
    R(\lambda_e)\propto  B(\lambda_e)^2\times
\begin{cases}
\lambda_e(\rhok^0(\lambda_i\pi/\lambda_e^2))^2,\quad \lambda_e\ll\lambda_e^{(c)}\\
\lambda_e,\quad\quad\qquad \lambda_e^{(c)}\ll\lambda_e\ll L,
\label{eq:response}
\end{cases}
\end{equation}
where 
\[
B(\lambda_e)=\frac{1}{1+\lambda_e^2/(2\pi\lambda_D)^2}
\]
is the response function containing the effect of the interactions, and $\hat{\rho}^0$ is the Fourier transform of the density profile of the cloud. In the experiments, we use a gaussian probe beam smaller than the cloud, in order to control the boundary effects in the transverse direction: hence the cloud's density profile is effectively limited in the transverse direction by $w$, the waist of the probe beam; $w$ is chosen significantly smaller than the cloud's size, and much larger than the modulation wavelength. In the longitudinal direction, we cannot avoid boundary effects, and accordingly, the diffracted intensity in the Bragg regime explicitly depends on the density profile of the cloud. In practice and to compare with the experiments, we have used expression \eqref{eq:app:density} for $\rho^0$.\\
iii) In the sub-Debye Raman-Nath regime $\lambda_e^{c}<\lambda_e<\lambda_D$, we then expect to see a response $R(\lambda_e) \propto \lambda_e$, whereas in the Raman-Nath regime for $\lambda_e > \lambda_D$, we expect to see $R(\lambda_e)$ \emph{decreasing}
with $\lambda_e$, ultimately as $\lambda_e^{-3}$: this is an effect of the interparticle repulsion. Our strategy is to look for this decreasing region in the experiment, in order to estimate $\lambda_D$.

\subsubsection{Comparison between experiment and theory}

\begin{figure}[t!]
\begin{center}
\includegraphics[width=0.45\textwidth]{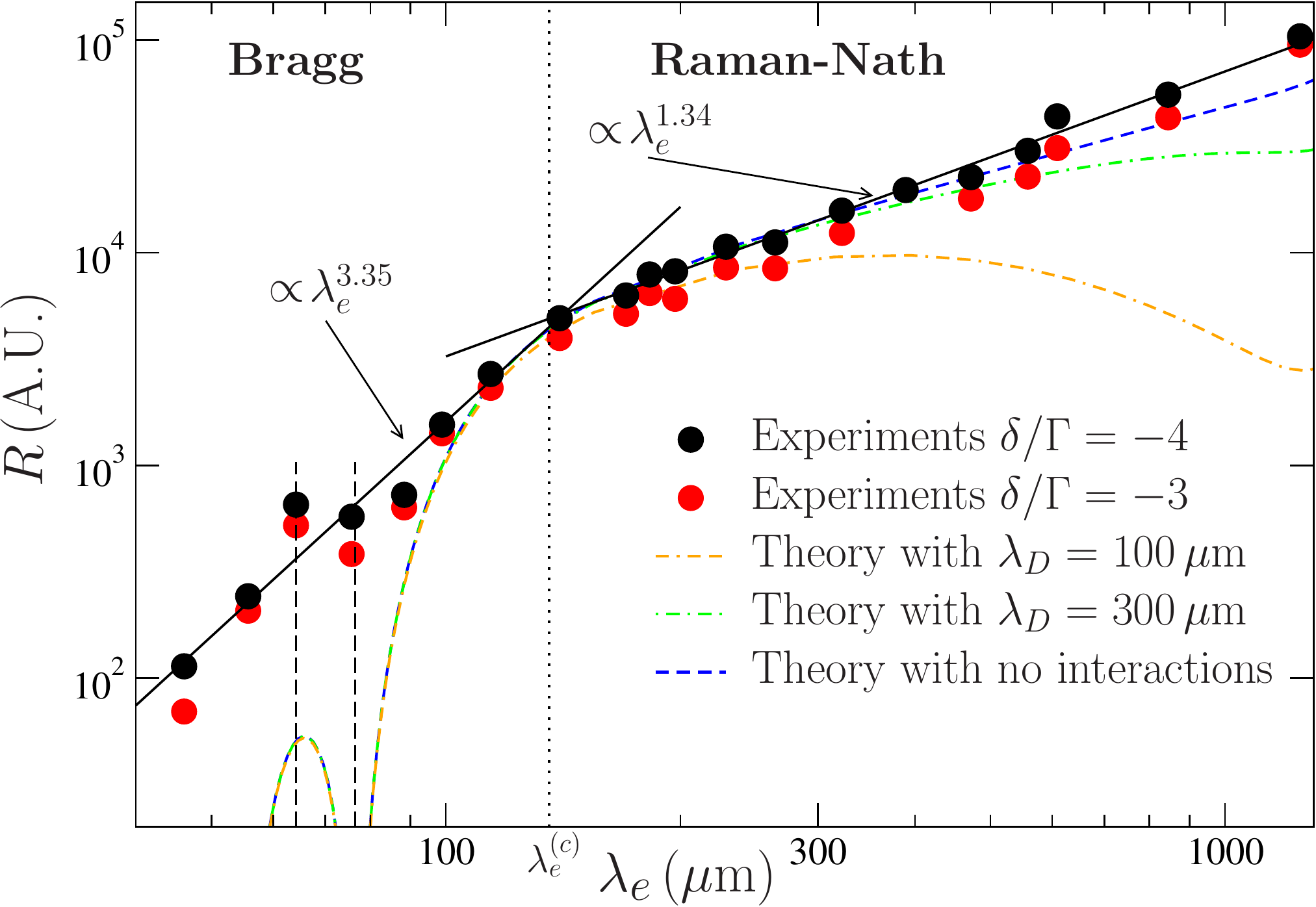}
\caption{Comparison of the total diffracted power $R(\lambda_e)$ in the experiment (red and black dots) and theory (lines). The detuning is $\delta/\Gamma=-3,-4$, $N\sim 10^{11}$, $w=2.2\,$mm. The theoretical curves use $w=2.2$mm, and $L=7.41$mm, which is the value extracted from Fig.\ref{fig:ocp:dxx} for $\delta/\Gamma=-4$; they are computed with Debye length $\lambda_D=100, 300\,\mu$m. The steepness $l$ of the step function in \eqref{eq:app:density} is chosen to be $l=1$mm (the theoretical curve only weakly depends on $l$).
We also show the theoretical limit case with no interactions $B(\lambda_e)=1$. 
The vertical dotted line indicates the theoretical position of the Bragg/Raman-Nath cross-over $\lambda_e^{(c)}=136\,\mu$m. The corresponding experimental value $\lambda_e^{(c),\rm{exp}}=142\,\mu$m is obtained at the intersection of the fitted  experimental data (for $\delta/\Gamma=-4$) in the Bragg $\propto\lambda_e^{3.35}$ and Raman-Nath region $\propto\lambda_e^{1.34}$. This latter exponent is not far ($1.34\simeq 1$) from the prediction of \eqref{eq:response} in the sub-Debye Raman-Nath regime \underline{without interactions}. The exponent in the Bragg regime depends on the specific details of the real experimental profile. The vertical dashed lines indicate a local maximum and a local minimum of the response in the Bragg regime, see Appendix \ref{sec:ocp:split}.}
\label{fig:ocp:diff_exp_3}
\end{center}
\end{figure}

We now analyze the experimental results using the above theory.
In Figure~\ref{fig:ocp:diff_exp_3} we plot the result of an experiment for a detuning $\delta=-3\Gamma$. We compare these results with the theoretical diffraction response of the profile~\eqref{eq:app:density}. The parameters $L, w, N$ are chosen to be the same as in the experiment. 
Indeed, the waist $w$ and atom number $N$ are well controlled and the size of the cloud $L$ can be extracted from a density profile. The smoothing length $l$  appearing in \eqref{eq:app:density} is chosen in the range suggested by the density profiles, see Fig.~\ref{fig:ocp:dxx}, and does not have much influence on the results.
The only adjusted parameter here is the vertical amplitude of the theoretical response (in arbitrary units), that we set so it coincides with the experimental curves.
The three theoretical curves correspond to three values for the Debye length $\lambda_D$: this modifies the response \eqref{eq:response}. 

The conclusions of this comparison are
\begin{itemize}
\item The Bragg/Raman-Nath crossover predicted in~\eqref{eq:ocp:criteria} is observed in the experiment, at the predicted location. 
\item In the Bragg regime the theoretical response is smaller than what is observed. In this region, the response is sensitive to the details of the density profile, and our simple assumption \eqref{eq:app:density} may not be good enough.
\item The theoretical analysis predict oscillations in the Bragg regime. While these oscillations are not clearly resolved in the experiments, some hints are visible on figure \ref{fig:ocp:diff_exp_3} (vertical dashed lines around $\lambda_e=70\,\mu$m). In Appendix \ref{sec:ocp:split}, we analyze in more details the theoretical and experimental diffraction profiles, to confirm that
the experimental observations are indeed a remnant of the theoretically predicted oscillations.
\item In the Raman-Nath regime close to the crossover, the slopes of experiment and theory are both about 1. For larger modulation wavelength, we expect the long-range effects to take place. We indeed see clearly on the theoretical curve with $\lambda_D=100\,\mu$m a decreasing response.
For $\lambda_D=300\,\mu$m this decrease occurs for larger $\lambda_e$ and is thus barely visible. For comparison, we plot (blue dashed line) the limit   
$\lambda_D\to \infty$, corresponding to a non interacting case.
The experimental data show no decrease for large wavelength: hence they are close to the "no interaction" case. More precisely, these data match the Coulomb predictions only if the Debye length is larger than $\sim 400\,\mu$m. 
Unfortunately, probing larger $\lambda_e$ is difficult and would be hampered by strong finite size effects.

\item In principle, from the analysis of the variations of $R$ with $\lambda_e$ in the Raman-Nath regime and for $\lambda_e \gg \lambda_D$, we could hope to test the validity of the $1/r^2$ force: this Coulomb model predicts a $-3$ exponent. However, this  $\lambda_e \gg \lambda_D$ regime is not seen in the experiments, and unfortunately the regime which is seen, $\lambda_e < \lambda_D$, is precisely the one where $R$ contains no signature of the interactions.

\end{itemize} 

\section{Conclusion}
\label{sec:conclusion}

We have proposed in this paper to use the response to an external modulation as an indirect way to measure the correlations inside the atomic cloud, and more generally to probe the effective interactions induced by the multiple photon scattering in large MOTs.

The modulation experiments and comparison with simulations did not show any evidence for a Debye length within the explored range, which could indicate a larger than expected value for $\lambda_D$ of at least $400~\mu$m for a detuning $\bar{\delta}=-4$. This seems consistent with direct numerical fits of the cloud's density profile, which suggest a Debye length as large as $1$ mm. Accordingly, an extension of the modulation experiment to larger wavelengths could be envisioned.
These values should be compared to the rough a priori estimate $\lambda_D \sim 100~\mu$m, based on the Coulomb model for the interaction between atoms and the observed size of the cloud. 
A clear theoretical explanation for the discrepancy between the a priori estimate for $\lambda_D$ and the bounds provided by the experiments is lacking. It is possible that the Coulomb model for the effective interactions between atoms reaches its limits in such large MOTs: the Coulomb approximation relies on a small optical depth, whereas it is around $1$ in experiments; or the spatial dependencies of the scattering sections may have to be considered. In either case, a refined model taking these effects into account would be considerably more complicated. It might also be that another mechanism controlling the maximum density, and hence the size of the cloud, is at play beyond multiple diffusion.

\appendix
\section{Linear response computations for the modulation experiment}
\label{sec:appA}

Writing the new density profile as a perturbation around the constant density $\rho_c$, $\rho(\r)=\rho_c+\delta\rho(\r)$, we can compute $\delta \rho$ at linear order 
using Eqs. \eqref{sf}, \eqref{density-ft}, 
\eqref{ocp-sf} and \eqref{eq:modulation} (this neglects the effect of the cloud's boundary): 
\begin{equation}
\delta\rho(x,y,z)=\dfrac{A}{\kb T}\rho_c B(\lambda_e)\sin(k_e x)
\label{eq:ocp:delta_rho}
\end{equation}
where 
\[ 
B(\lambda_e)=\dfrac{1}{1+\lambda_e^2/(4\pi^2 \lambda_D^2)}~,~\lambda_e=\frac{2\pi}{k_e}
\]
and $A$ is the small amplitude of the modulating potential.
Hence the modulated profile has a clear amplitude dependence on the modulation wavelength $\lambda_e$ and it is characteristic of Coulomb interactions (another force would have given a different result). When the modulation wavelength is increased beyond the Debye length ($L>\lambda_e>\lambda_D$), the response decreases, which means that large scale inhomogeneities are more difficult to create: this is an effect of repulsive long range interactions. Therefore, measuring this response function should provide information on the interactions inside the cloud.

The density modulation of the cloud is measured by diffraction: the diffracted amplitude at wavelength$\lambda_e$ is related to the response function $B(k_e)$. However, this relationship is not straightforward. In particular, we shall see now that there are two distinct diffraction regimes, Bragg at small wavelength, and Raman-Nath at large wavelength.

The diffraction profile is proportional to the structure factor, which is for the modulated cloud, using the definition~\eqref{eq:structure_factor}:  
\begin{equation}
S(\vec{k})=S^0(\vec{k})+\dfrac{2}{N}\delta\rhok(\k)\rhok^0(\k)+\delta\rhok(\k)^2+\O{\text{correlation}},
\label{eq:ocp:s_exact}
\end{equation}
where $S^0$, $\rhok^0$ are respectively the structure factor and the Fourier transform of the effective cloud's profile without external modulation; note that it actually corresponds to the cloud's profile truncated in the $x$ and $y$ direction by the gaussian probe beam. Hence here $N$ corresponds to the number of diffracted atoms, ie within the gaussian probe beam. We will neglect the correlations because they are very small as we have seen in section \ref{sec:diffraction}.
The Fourier transform of the modulated  cloud $\delta\rhok(\k)$ can be related to the Fourier transform of the unperturbed cloud $\rhok^0(\k)$, taking into account the shift in $\k$ induced by the $\sin (k_e x)$ function $k_x\to k_x\pm k_e$.
The diffracted peaks correspond to maxima of the structure factor and are situated around the wavenumber $|\k|\simeq |\k_e|$. To compute their amplitude and shape one can expand in \eqref{eq:ocp:s_exact} around $k=k_e$, and $\phi_k=0$ or $\pi$ (these two angles correspond experimentally to the two diffraction peaks observed, see Fig.~\ref{fig:angles} for definition of $k$ and $\phi_k$).\\

We probe a wavenumber region $k_e\in[\sim 10^3,\sim 10^5]$\,m$^{-1}$, with $k_i=2\pi\frac{10^6}{0.78}$\,m$^{-1}$, so that $k_e/k_i\ll 1$. This justifies the following expansion
\begin{equation}
\begin{split}
|k_e\,\vec{e}_k- k_e\,\vec{e}_x|&= \dfrac{k_e^2}{2 k_i}+k_e\times \O{\left (\dfrac{k_e}{2 k_i}\right )^2}
\\&\simeq k_z\neq 0.
\end{split}
\label{eq:k_dev}
\end{equation}
In the perturbed density profile, it yields at the diffracted peak $k\simeq k_e$
\begin{equation}
\rhok(k_e)\simeq\rhok^0(k_e)-\dfrac{A}{2 \kb T}B(k_e)\left (\rhok^0\left (2 k_e\right )-\rhok^0\left (\dfrac{k_e^2}{2 k_i}\right )\right ).
\label{eq:ocp:tf_pert}
\end{equation}
Since $\rhok(k=0)=N$ and the Fourier transform of the profile decreases very quickly to $0$ with increasing $k$ (the more regular $\rho(r)$ is, the faster its Fourier transform goes to $0$) the dominant term in \eqref{eq:ocp:tf_pert} is the last one, provided $NA/(\kb T) \gg 1$ (this is typically the case in experiments) and $k_e\gtrsim 1/L$. Hence the diffracted peak maximum intensity is given by
\begin{equation}
S(k_e)\simeq 1+\dfrac{1}{N}\left (\dfrac{A}{2 \kb T}\right )^2B^2(k_e) (\rhok^0(k_z))^2.
\label{eq:ocp:S_dom}
\end{equation}
Thus the diffraction response depends on the longitudinal density profile and not only on the response function $B(k_e)$. The density dependence crossovers at $k_z L\sim 1$, which defines a critical modulation wavelength $\lambda_e^{(c)}$ (or wavenumber $k_e^{(c)}$)
\begin{equation}
\lambda_e^{(c)}=2\pi\sqrt{\dfrac{L}{2k_i}}=\sqrt{\pi L\lambda_i}\quad\text{or}\quad k_e^{(c)}=\sqrt{\dfrac{2k_i}{L}}.
\label{eq:ocp:criteria_app}
\end{equation}
It separates on one side the Raman-Nath regime $k_z L\ll 1$, where the diffracted peak intensity depends only on the response function, and on the other side the Bragg regime $k_z L\gtrsim 1$, where $\hat{\rho}^0(k_z)$ is not constant and decreases quickly to zero. Thus in this latter regime there is an additional dependence related to the Fourier transform of the density profile, that we call ``density effect".
Note that in the context of ultrasonic light diffraction this criterion~\eqref{eq:ocp:criteria} separating Bragg and Raman-Nath regimes is also known~\cite{klein_unified_1967}.
For a cloud of radius $L\thickapprox 6\,$mm and a laser $\lambda_i\simeq \lambda_L=780$\,nm, the crossover is expected around 
$\lambda_e^{(c)}\thickapprox 120\,\mu$m. 

It must also be noted that the experimentally measured quantity is not the peak amplitude $S(k_e)$, but rather the diffracted power $R(k_e)$:
this brings an extra dependence on $k_e$. To simply show this, one can expand the structure factor around the peak and, assuming for instance a Gaussian shape around the maximum, deduce a linear dependence on the modulation wavelength $\lambda_e=2\pi/k_e$ (the precise form of the shape around the maximum does not modify this linear dependence). To summarize, we expect to measure 
\begin{equation}
R(k_e)\propto B^2(k_e)\times
\begin{cases}
\lambda_e(\rhok^0(\lambda_i\pi/\lambda_e^2))^2,\quad \lambda_e\ll\lambda_e^{(c)}\\
\lambda_e,\quad\quad\qquad \lambda_e^{(c)}\ll\lambda_e\ll L.
\label{eq:response_app}
\end{cases}
\end{equation}
In this expression, both the density dependence and response function $B(k_e)$ are a priori unknown. In order to obtain a well defined theoretical prediction, we 
assume for the cloud's profile a symmetrized Fermi function \cite{sprung_symmetrized_1997}, ie a step smoothed over a length scale $l$. In the direction perpendicular to the probing beam, the cloud is effectively limited by the waist of the probing laser $w$; we assume a gaussian laser profile. 
This yields a simplified effective density profile 
\begin{equation}
\rho^0(r_\perp,z)\propto\dfrac{l}{L}\dfrac{ \sinh \left(\frac{L}{l}\right)}{\cosh \left(\frac{L}{l }\right)+\cosh \left(\frac{z}{l }\right)}\exp \left(-\frac{2 r_\perp^2}{w^2}\right).
\label{eq:app:density}
\end{equation}
Its associated structure factor can be evaluated analytically thanks to~\cite{sprung_symmetrized_1997}. 
Putting together all the results of this section, we obtain the theoretical predictions shown on Fig.\ref{fig:ocp:diff_exp_3}.

\section{Oscillations in the Bragg regime}
\label{sec:ocp:split}
\begin{figure}[!htbp]
	\begin{center}
		 \subfigure[$\lambda_e=64.2\,\mu$m]
		 	{
			\includegraphics[width=0.22\textwidth]{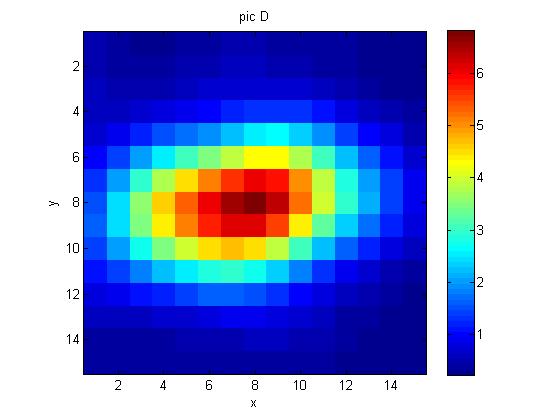}\label{fig:ocp:exp_tache_simple}
             }
		 \subfigure[$\lambda_e=75.7\,\mu$m]
		 	{
			\includegraphics[width=0.22\textwidth]{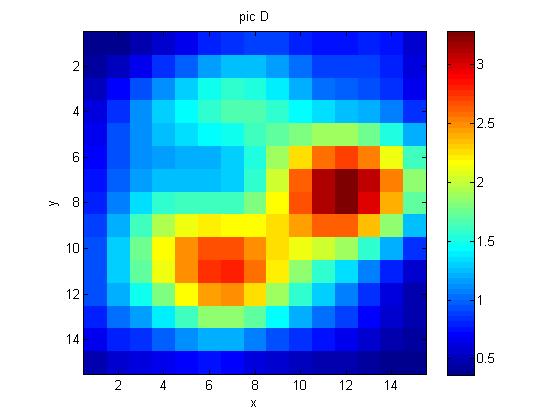}\label{fig:ocp:exp_tache_double}
             }      
             \subfigure[$\lambda_e=64.2\,\mu$m]
		 	{
            \label{fig:ocp:m_tache_simple}
            \includegraphics[width=0.20\textwidth]{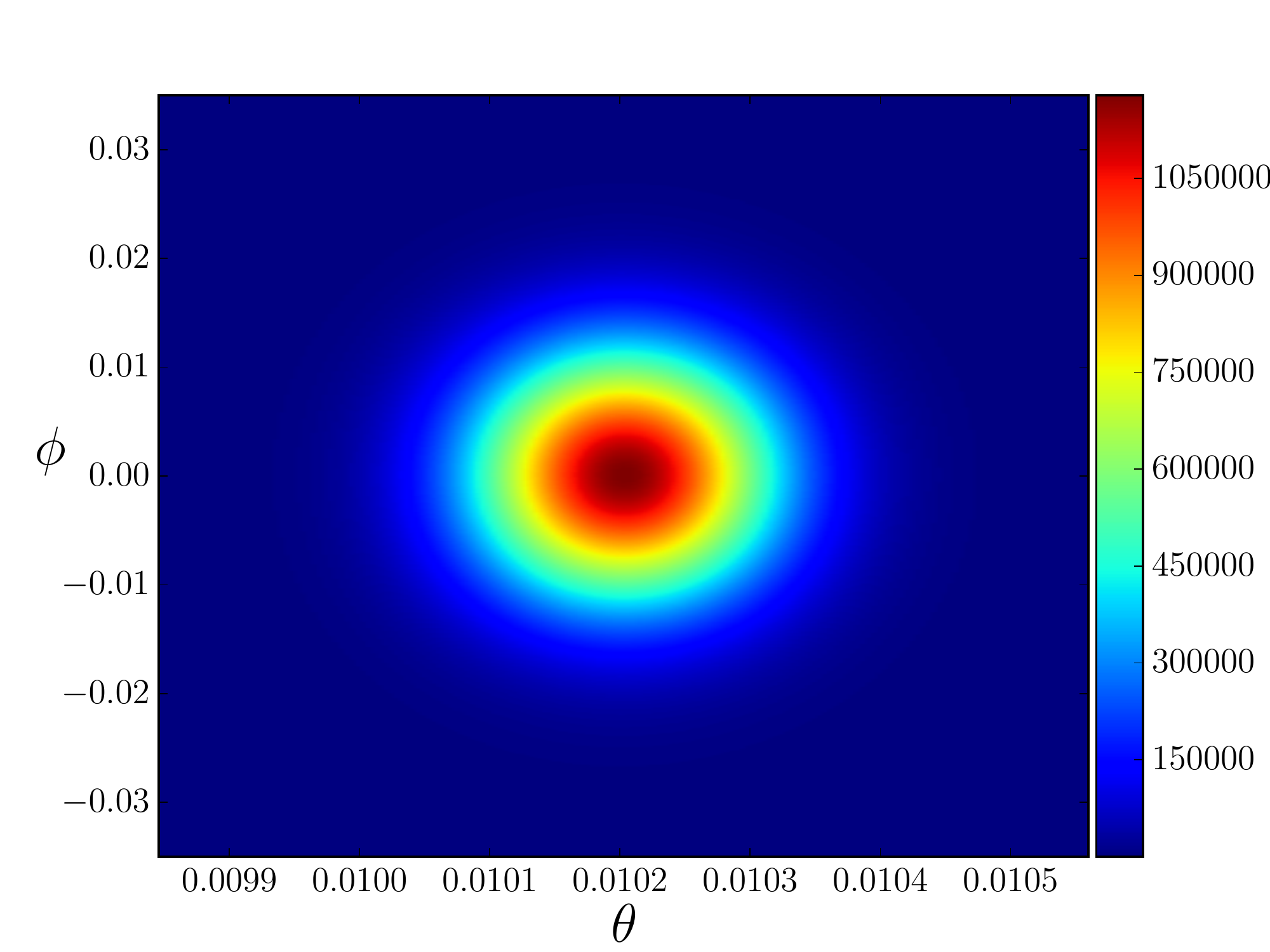}\label{fig:ocp:tache_simple}
             }
		 \subfigure[$\lambda_e=76.5\,\mu$m]
		 	{
            \label{fig:ocp:m_tache_double}
            \includegraphics[width=0.20\textwidth]{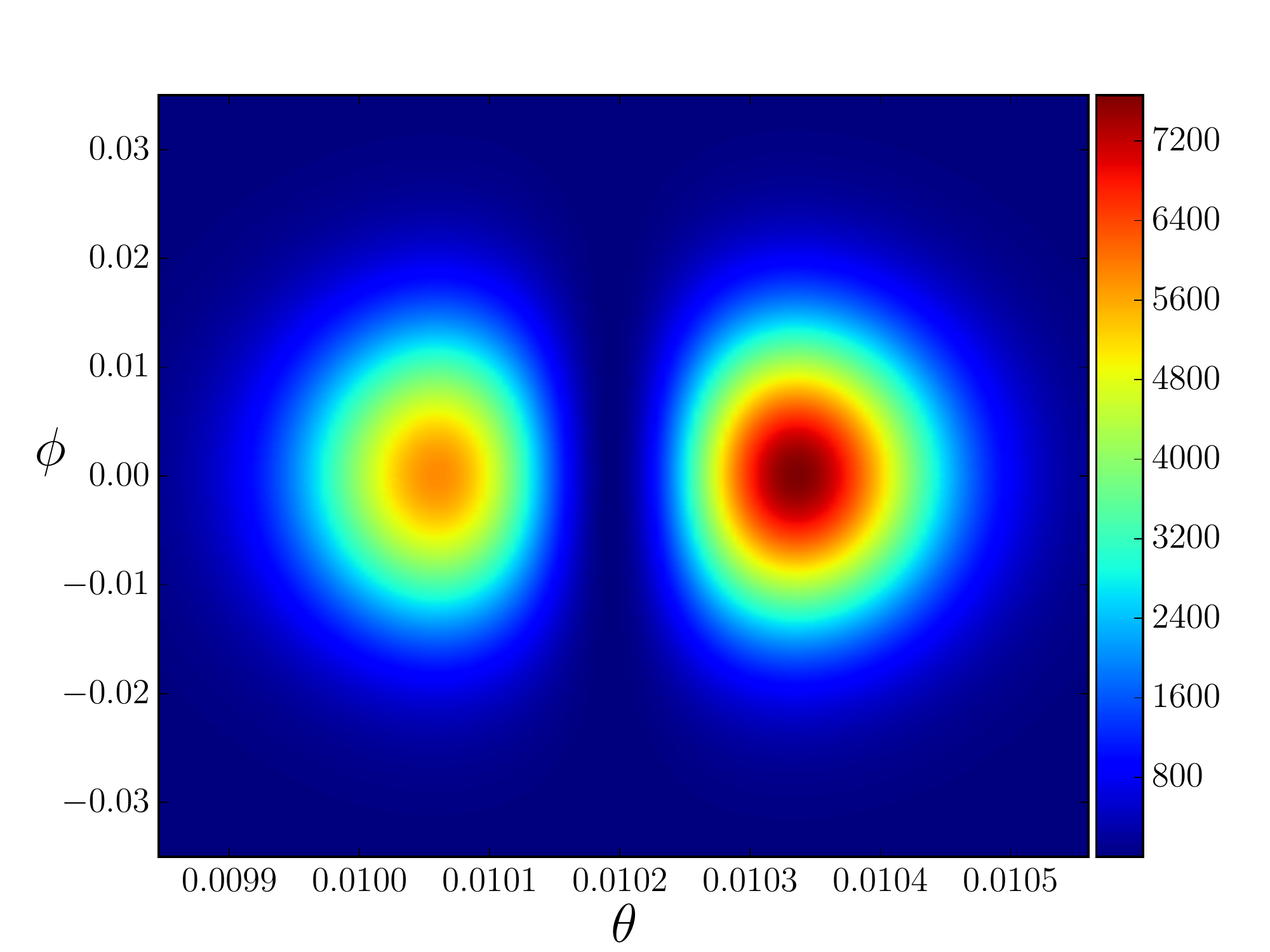}\label{fig:ocp:tache_double}
             }             
	\end{center}
	\caption{Experimental (top) and theoretical (bottom) diffracted beams for $\lambda_e=64.2$ and $75.68\,\mu$m.} 	
\label{fig:ocp:exp}
\end{figure}

In the Bragg regime, the shape of the diffracted beams observed in the experiment shows some variations, as seen on Figure~\ref{fig:ocp:exp_tache_double}: for $\lambda_e=75.7\mu m$, the diffracted beam is split in two; this corresponds to the right dashed vertical line in Fig.~\ref{fig:ocp:diff_exp_3}.
Can we explain this observation? One has to remember that the response depends on the longitudinal profile~\eqref{eq:ocp:tf_pert}; thus around a peak $k=k_e+\delta k$, the response is
\[
S(k)\propto S^0\left (\dfrac{k_e^2+2k_e\delta k}{2k_i}\right ).
\]
$S^0(k)$ is the Fourier transform of the effective density profile \eqref{eq:app:density}. In the $z$-direction, this profile is a smoothed step, and this induces oscillations in its Fourier transform and in $S^0$; the locations of the local minima and maxima of these oscillations mainly depend on the cloud's size $L$, and only very weakly on the details of \eqref{eq:app:density}, such as the smoothing length scale~$l$. If $k_e^2/(2k_i)$ happens to correspond to a local minimum of $S^0$,  the diffracted beam can be split in two. 

We illustrate this with our theoretical model \eqref{eq:app:density}, with parameters $L$ and $w$ provided by the experiments, and $l$ chosen to be $1$mm (the results depend very weakly on $l$).
Figure~\ref{fig:ocp:tache_double} shows the theoretical diffracted beam for $\lambda_e=76.5\,\mu$m, where splitting 
occurs: this value of $\lambda_e$ is very close 
to the one for which splitting is indeed experimentally observed.
In Figure~\ref{fig:ocp:exp_tache_simple} we show an experimental image for $\lambda_e=64.2\,\mu$m (this corresponds to the left vertical dashed line of Figure~\ref{fig:ocp:diff_exp_3}) where no splitting occurs. The theoretical prediction  Fig.~\ref{fig:ocp:tache_simple} indeed does not show any splitting.

This analysis provides a satisfactory explanation of the experimental observation, and suggests that the Bragg regime is well understood.
These features have unfortunately nothing to do with the Debye length we are looking for: they are related to the global cloud's shape.

%%%%%%%%%%%%%%%%%%%%%%%%%%%%%%%%%%%%%%%%%%%%%%%%%%%%
%%%%%%%%%%%%%%%%%%%  Bibliography %%%%%%%%%%%%%%%%%%
%%%%%%%%%%%%%%%%%%%%%%%%%%%%%%%%%%%%%%%%%%%%%%%%%%%%

\end{document}